\documentclass[aps,prl,reprint,groupedaddress,showpacs]{revtex4-1}

\usepackage{graphicx}
\usepackage{dcolumn}
\usepackage{bm}
\usepackage{amssymb}
\usepackage{mathrsfs}

\begin{document}

\title{Experimental Study of Quantum Noise in Optical Heterodyne Detection
}


\author{Dechao He}
\author{Boya Xie}
\author{Yu Xiao}
\author{Sheng Feng}
\email[]{fengsf2a@hust.edu.cn}
\affiliation{
MOE Key Laboratory of Fundamental Quantities Measurement, School of Physics, Huazhong University of Science and Technology, Wuhan 430074, China}


\date{\today}

\begin{abstract}
We experimentally investigate the quantum-noise performance of a conventional heterodyne detector and find significant discrepancy between experiment and theory. Further investigations are highly recommended for deeper insight into the physics related to the quantum noise in optical heterodyne detection.
\end{abstract}
 

\maketitle


As a critical facet in quantum mechanics, the uncertainty principle sets a fundamental limit to the precision with which two complementary variables of a quantum object can be measured simultaneously. In a recent revival of interest in this principle, it was extended in terms of entropy to the case where quantum entanglement is involved \cite{berta2010,li2011,prevedel2011}. According to these studies, the outcomes of two non-commuting observables of a quantum object may be predicted precisely if assisted by entanglement \cite{berta2010,li2011}. In other words, the uncertainty in simultaneous measurement of two conjugate variables can be reduced down to zero in the presence of entanglement \cite{li2011,prevedel2011}. Here we present an experiment in which two conjugate quadratures of an electromagnetic (e.m.) field were simultaneously measured with an optical heterodyne detector without assistance of entanglement, with experimental results that challenge existing theory.

To measure the quadratures of an e.m. field, one usually mixes the studied field with a stronger reference field (namely, local oscillator) and measure the intensity of the mixed fields, with the intensity proportional to a quadrature of the measured field depending on the relative phase between the two fields \cite{yuen1980,yuen1983}. Such kind of measurement is referred to as optical homodyning (heterodyning) when the difference of the optical frequencies of the two mixed fields is zero (nonzero). Homodyne detectors are phase-sensitive and free of quantum noise \cite{caves1982}. They have been intensively exploited to measure non-classical states of light \cite{slusher1985,Breitenbach1997,lvovsky2001} and for quantum information experiments \cite{li2002,bowen2003,weedbrook2012}. All experiments agree with theory.

\begin{figure}
\includegraphics[scale=0.35]{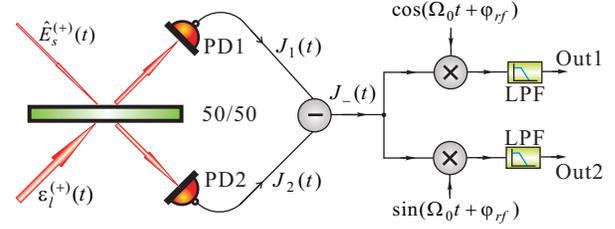}
\caption{\label{fig:scheme1} (color online) Proposed scheme for study of the quantum noise in optical heterodyne detection.  At a balanced (50/50) beamsplitter, a coherent signal light $\hat{E}_s^{(+)}(t)$ is mixed with an optical local oscillator $\mathscr{E}_l^{(+)}(t)$. The mixed light at each output port of the beamsplitter is collected by a photodetector (PD1 or PD2) and the differenced photocurrents $\hat{J}_-(t)\equiv J_1(t)-J_2(t)$ are post-processed for simultaneous measurement of conjugate quadratures of the signal field $\hat{E}_s^{(+)}(t)$ by using separate radio-frequency (r.f.) oscillators with 90$^{\circ}$ phase difference \cite{shapiro1984,yamamoto1986}. $\hat{E}_s^{(+)}(t)$ is excited at angular frequency $\omega_s$ and $\mathscr{E}_l^{(+)}(t)$ at $\omega_l$ ($\omega_l-\omega_s\equiv\Omega_0\ne 0$). $\cos(\Omega_0 t+\varphi_{rf})$ and $\sin(\Omega_0 t+\varphi_{rf})$ represent the two r.f. oscillators, respectively. LPF: Low-pass filter.
}
\end{figure}

On the contrary, research on the quantum noise in optical heterodyne detection was mainly limited to theoretical exploration \cite{oliver1961,haus1962a,personick1971,yuen1980,yuen1983,shapiro1984,yamamoto1986,collett1987,caves1994,ou1995,buonanno2003,rubin2007,shapiro2009} with very few experimental results reported \cite{meers1991,lixiao1999}. Specifically, conforming to the uncertainty principle, a 3 dB extra quantum noise in optical heterodyning, as a result of simultaneous measurement of conjugate quadratures of an e.m. field, has been theoretically predicted for a long time \cite{yuen1980,shapiro1984,yamamoto1986,caves1994,buonanno2003,shapiro2009}. The recent interest in the uncertainty principle \cite{berta2010,li2011,prevedel2011,ozawa2003,rozema2012} naturally urges an experimental realization of the original proposal (Fig. \ref{fig:scheme1}) to test out the relevant theories. Besides its great importance from the viewpoint of fundamental research, experimental investigation on the quantum noise in optical heterodyning also finds itself very useful in practical applications \cite{buonanno2003}. The predicted 3 dB extra quantum noise, if exists, makes heterodyning less competitive than homodyning in precision measurements. This could play a crucial role in the design of advanced experiments for precision measurements, such as next-generation LIGO experiment \cite{buonanno2003,ligo2013}. 

\begin{figure*}
\includegraphics[scale=0.40]{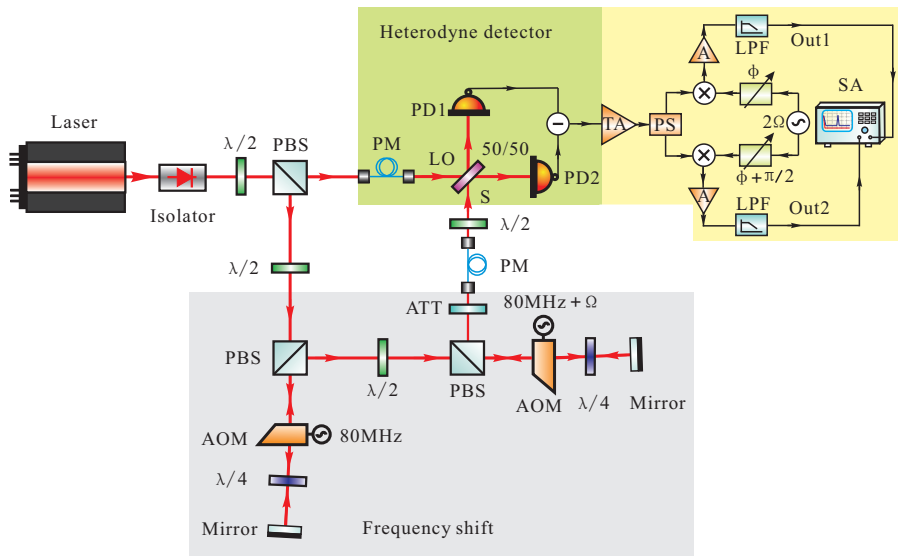}
\caption{\label{fig:scheme2}(color online) Experimental setup for study of the quantum noise in optical heterodyne detection. $\lambda/2$: Half-wave plate. PBS: Polarizing beamsplitter. AOM: Acousto-optic modulator. $\lambda/4$: Quarter-wave plate. ATT: Optical power attenuator. PM: Single-mode polarization-maintaining fiber, used as spatial-mode cleaner. S: Signal light. LO: Optical local oscillator. PD1 \& PD2: Photodiodes. TA: Transimpedance amplifier followed by an electric power amplifier. PS: Power splitter. A: Electric power amplifier. $2\Omega$: r.f. signal generator. $\phi$ and $\phi+\pi/2$: Electric phase shifters. LPF: Low-pass filter. SA: Dual-channel FFT spectrum analyzer. The frequency offset $\Omega$ in the electric signal fed into the driver of the second AOM was set to be 0.4 MHz for optical heterodyning or zero for optical homodyning.
}
\end{figure*}

The research activities on the quantum noise in optical heterodyning can be traced back to early 1960's, when Oliver showed \cite{oliver1961} that the signal-to-noise ratio (SNR) with heterodyning is two-fold worse than that with homodyning. Before long, Haus and Townes came up with a similar conclusion by connecting Oliver's treatment to the uncertainty principle \cite{haus1962a}. However, both Oliver and Haus carried out their detailed calculations based on classical physics. In 1971, Personick presented an image-band interpretation of the optical heterodyne noise \cite{personick1971}, arguing that it is the image-band modes that make optical heterodyning quantum-mechanically twice noisier than homodyning. Using Personick's concept of image-band mode, Yuen and Shapiro revisited the same subject with more rigorous theoretical developments \cite{yuen1980}, leading to a proposal for simultaneous measurement of both conjugate quadratures of an e.m. field with a optical heterodyne detector and an explanation of how the uncertainty principle plays its role in introducing 3 dB extra quantum noise into optical heterodyning \cite{shapiro1984,yamamoto1986,caves1994}. 
It is this extra quantum noise in heterodyning that the current work focuses on and, by experimentally detecting this noise, one wishes to put previous theoretical developments, especially those related to the uncertainty principle, under experimental test.

In the experimental realization of the proposal depicted in Fig. \ref{fig:scheme1}, an essential work was to assure that the quantum noise in the experiment dominated over all classical noises. Howbeit this is not difficult in homodyning experiments, one should bear in mind that, according to the original proposal \cite{shapiro1984,yamamoto1986,buonanno2003}, the detector-generated photocurrents must be post-processed with electronics devices, which inevitably introduce classical noises that may submerge the quantum noise under study. The success of the experiment, the detailed description of which is given in Fig. \ref{fig:scheme2}, relied heavily on the extra-low-noise electronics developed in lab.

A continuous-wave single-frequency coherent light beam (spectral linewidth $<1$ kHz for 0.1 s  measurement time, $\lambda=$ 1064 nm) from a laser (Mephisto, Innolight GmbH) was split into two, one of which served as local oscillator with the other sent through two AOM's (Crystal Technology, LLC) for frequency shifting (upshift in one AOM and downshift in the other). The frequency-shifted beam was used as an input signal for heterodyning (heterodyne frequency $2\Omega=0.8$ MHz) at a balanced (50/50) optical beamsplitter. 
Two photodiodes (ETX 500, JDS Uniphase) collected the light from the output ports of the beamsplitter and the differenced photocurrents 
were post-processed for the experiment of simultaneous measurement of conjugate quadratures of the signal light field, as follows \cite{shapiro1984,caves1994}: 

Differentiated photocurrents were first processed by a transimpedance amplifier (feedback resistor 5 k$\Omega$) and a power amplifier (Mini-circuits, ZFL-500LN+) in series. The electric time-varying signals were then equally split with a power splitter (Mini-circuits, Z99SC-62-S+) into two parts, each of which was mixed at a mixer (Mini-circuits, ZP-3+) with separate r.f. oscillators (generated by an Agilent dual-channel function generator, 33522A) with 90$^{\circ}$ phase difference, to obtain two electric signals, each carrying quantum-noise information of one of two conjugate quadratures of the signal field \cite{shapiro1984,yamamoto1986,buonanno2003}. Since the signal light was in a coherent state with phase-insensitive quantum noise, the global phase of the two r.f. oscillators did not have to be controlled for the purpose of this experiment. The predicted 3 dB extra quantum noise in heterodyning, if existed, should be present in the output signals of the two mixers \cite{shapiro1984,caves1994}. Both signals, too weak for shot-noise-limited record, were yonder amplified by about 40 dB in power with home-made amplifiers before final data record was finished with a dual-channel FFT spectrum analyzer (SR785, Stanford Research Systems).

\begin{figure}
\includegraphics[scale=0.28]{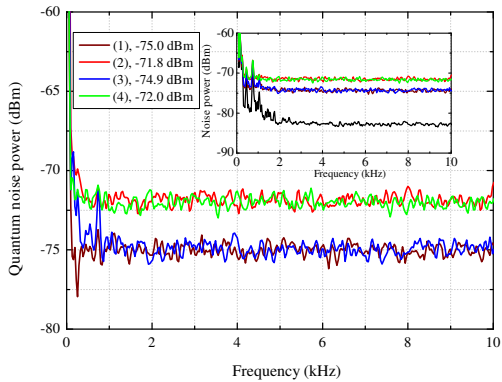}
\caption{\label{fig:data1}(color online) Quantum-noise floors (traces (1)-(2), dark-current noise subtracted) of the heterodyne detector when one of the r.f. oscillators was removed from the experimental setup and, thereby, only one quadrature of the signal field was measured. The power of the optical oscillator was 1.45$\pm$0.03 mW for trace (1) and 2.90$\pm$0.06 mW for trace (2), respectively. The statistical errors were $\pm$0.3 dBm for both traces. Traces (3) and (4) were quantum-noise floors in optical homodyning, with optical-oscillator powers similar to those for traces (1) and (2), respectively. 
RBW = 32 Hz. Number of r.m.s. averages was 100. Inset: Noise power spectra corresponding to traces (1)-(4) before dark-current noise subtractions were performed, shown together with a typical dark-current noise power spectrum (black curve). 
}
\end{figure}

Before the experiment got started, one must demonstrate the capability of the experimental setup to detect the quantum-noise floors of light at appropriate power levels. To this end, one compared the power levels of the detected noises with the theoretical expectations. The observed noise-power density was -90.1$\pm$0.3 dBm/Hz for a 1.45$\pm$0.03 mW optical oscillator, in which case the theoretical expectation for the quantum-noise power density was -88.8$^{+1.1}_{-1.2}$ dBm/Hz with the uncertainty determined by the errors in the measurement of the experimental parameters used in the calculation. Similar result was obtained for a 2.90$\pm$0.06 mW optical local oscillator. 
Moreover, one observed that doubling the power of the optical oscillator resulted in a 3.1$\pm$0.3 dB uptick (Fig. \ref{fig:data1}), which would be otherwise 6 dB if classical noises of light dominated, in the noise power level of light. 

\begin{figure}
\includegraphics[scale=0.28]{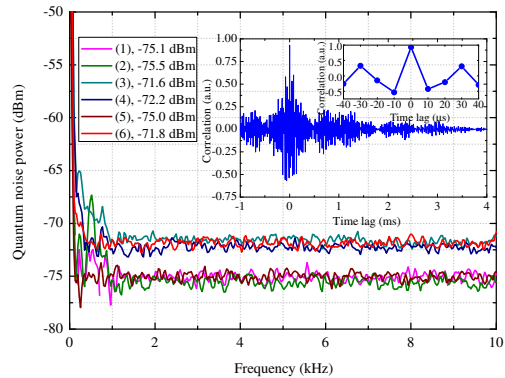}
\caption{\label{fig:data2}(color online) Power spectra (traces (1)-(4), dark-current noise subtracted) of the quantum noises of the heterodyne detector when simultaneous measurement of conjugate quadratures of the signal field was being performed. Trace (1) was the quantum-noise power spectrum for one quadrature and trace (2) for the conjugate component with $\pm$0.3 dBm statistical errors, with a 1.45$\pm$0.03 mW optical oscillator. The same is true for traces (3) and (4), except that the optical oscillator power was 2.90$\pm$0.06 mW. Traces (5)-(6) are nothing but traces (1)-(2) in Fig. \ref{fig:data1}, serving here as quantum-noise power level references for observation of the predicted extra quantum noise. RBW = 32 Hz. Number of r.m.s. averages was 100. Inset: The cross correlation of shot-noise-limited time-varying signals, each recorded by one of the two channels of the FFT spectrum analyzer when the phase difference of the two r.f. oscillators was set zero.
}
\end{figure}

The quantum-noise floors identified in the above procedure can at the same time serve as references for observing the predicted 3 dB extra quantum noise resulted from simultaneous measurement of conjugate quadratures of the signal field. The experiment was carried out following the proposed scheme \cite{shapiro1984,yamamoto1986,caves1994}, with results presented in Fig. \ref{fig:data2} where the predicted 3 dB extra quantum noise did not show up! 

Provided that all classical noises of light were negligible, the primary imperfections that might prevent one from observing the predicted extra quantum noise in the experiment were: Non-unity photon collection efficiency, non-perfect spatial-mode matching between the signal field and the optical oscillator, and asynchrony in the measurement of conjugate quadratures of the signal field. If the 3 dB extra quantum noise was present in optical heterodyning, the quantum-noise power level in simultaneous measurement of both conjugate quadratures should be $10\log_{10}\{\eta [1+\frac{(\delta\hat{X}_s)^2}{(\Delta\hat{X}_s)^2}]+(1-\eta)\frac{(\Delta\hat{X}_v)^2}{(\Delta\hat{X}_s)^2}\}$ in dB, where $\eta$ characterizing the effect of photon loss and spatial-mode mismatching, above the shot-noise floor. $\Delta\hat{X}_s$ and $\Delta\hat{X}_v$ are the quantum noise of the measured quadrature of the signal field $\hat{E}_s(t)$ and that of the vacuum $\hat{V}(t)$, respectively. $(\Delta\hat{X}_s)^2=(\Delta\hat{X}_v)^2$, since $\hat{E}_s(t)$ was in a coherent state in the experiment. $\delta\hat{X}_s$ is the disturbance on $\hat{X}_s(t)$ as required by the uncertainty principle. For simultaneous measurement of conjugate quadratures, $(\delta\hat{X}_s)^2 = (\Delta\hat{X}_s)^2$  and asynchrony in the measurement results in $(\delta\hat{X}_s)^2 \le (\Delta\hat{X}_s)^2$ \cite{ozawa2003}.

As a result of non-perfect photon-collection efficiency and spatial-mode mismatching between the signal light and the optical oscillator, the vacuum comes into play \cite{leon2005,yamamoto1986} and the aforementioned 3 dB extra quantum noise ($\eta=1$) could escape detection in optical heterodyning when $\eta\rightarrow 0$. In the experiment, the overall photon-collection efficiency was 67\%. The interference between the optical beams had a fringe with visibility of 97\%, measured with 100 $\mu$W light beams. One potential worry could be whether the visibility was significantly affected when one performed the heterodyning experiment with strongly attenuated signal light, the power of which was reduced down to 1 nW. However, the experiment was designed such that the attenuation of the signal light took place before it entered into the PM-fiber, which served as a spatial filter to keep the spatial profile of the beam unchanged during the experiment, ensuring reliable fringe visibility. 

As for the experimental requirement for the simultaneity in the measurement of conjugate quadratures of the signal field, it was mainly determined by the frequency bandwidth of the quantum noise of interest. If one was to study the quantum noise within wider (narrower) bands, the asynchrony in the measurement may be tolerated with a smaller (larger) temporal uncertainty. In the experiment, the data were collected for quantum noise from DC to 10 kHz, which demanded a measurement asynchrony of the order of 100 $\mu$s. To quantify the asynchrony in the quadrature measurement, one may measure the cross correlation of the shot-noise-limited time-varying signals produced at the two outputs of the setup, each recorded by one of the two channels of the FFT spectrum analyzer. The data presented in the inset in Fig. \ref{fig:data2} implies an asynchrony of $\pm$10 $\mu$s in the quadrature measurement, with this number limited by the maximal bandwidth of the spectrum analyzer. The actual asynchrony should be much better than 10 $\mu$s. Taking into account all the imperfections in the experiment, if the 3 dB additional quantum noise was present in optical heterodyning as predicted, one would expect to observe a 2.0 dB increment of quantum-noise power levels, which did not occur in the experiment, when both conjugate quadratures of the signal field were measured with $\pm$10 $\mu$s asynchrony. 

It has come to one's attention that the 3 dB extra quantum noise in heterodyne detection might be understood by someone in the following way \cite{haus1962a,haus2011}: The quantum-noise floor of a heterodyne detector is the same as that of a homodyne detector, but the photoelectric signal produced by the heterodyne detector is half of that by the homodyne detector, resulting in a two-fold reduction in SNR in heterodyne detection. If this interpretation of the 3 dB extra quantum noise is correct, then one must know at the same time both the photoelectric signal and the quantum-noise floor in the experiment to obtain conclusive results. Notwithstanding, the above interpretation does not agree with experiment, as shown in Table \ref{table1} where $P_s$ stands for the detected power of the signal-carrying light beam (taking into account the photon collection efficiency and spatial mode mismatching between the light beams). SNR$_{o}$=20log$_{10}\sqrt{N}$ is the theoretically expected SNR in power of an optical signal carried by coherent light, where $N$ is the photon number detected within certain period of time, which was $10^{-4}$ s in the experiment. SNR$_{e}$ is the SNR in power of the photoelectric signal generated by the heterodyne detector. It was measured by directly comparing the photoelectric signal power with the quantum-noise power displayed on the spectrum analyzer (Agilent, N9320B), with an uncertainty primarily determined by the statistical error of the quantum-noise power level.
\begin{table} 
\caption{\label{table1} SNR of an optical signal carried by coherent light and SNR of the corresponding photoelectric signal produced in heterodyne detection}
\begin{ruledtabular}
\begin{tabular}{c|c|c}
\ \ \ \ $P_{s} $ (nW) \ \ \ & SNR$_{o}$ (dB)\ \ \ \ \ \ & SNR$_{e}$ (dB) \ \ \ \ \ \ \\ \hline
0.9$\pm$0.1 & 56.6$\pm$0.6  \ \ \ \ \ \ & 56.9$\pm$0.8 \ \ \ \ \ \ \\\hline
2.0$\pm$0.1 & 60.4$\pm$0.2  \ \ \ \ \ \ & 61.5$\pm$0.8 \ \ \ \ \ \ \\\hline
6.0$\pm$0.1 & 65.2$\pm$0.1  \ \ \ \ \ \ & 66.0$\pm$0.8 \ \ \ \ \ \ \\
\end{tabular}
\end{ruledtabular}
\end{table}

At this point, on one hand, one would not claim that the uncertainty principle was broken in the optical heterodyning experiment, until a comprehensive study is carried out on the proposed scheme for quantum measurement of e.m.-field quadratures \cite{shapiro1984,caves1994} to exclude all implicit improper assumptions, if any, in the theoretical model. On the other hand, one should note that the 3 dB extra quantum noise has also been expected in optical heterodyne detection in Personick's image-band interpretation of heterodyne noise \cite{personick1971,yuen1983} and in the scenario of phase-insensitive linear amplifiers \cite{caves1982}, without directly invoking the uncertainty principle. Surely, it is out of the scope of this work to address all the problems relevant to the discrepancy between the experiment and theory. Rather than believing that this discrepancy resulted from mistakes in a variety of theoretical developments, one suspects some unknown physics at the quantum level in optical heterodyne detection.


To conclude, we have experimentally studied the quantum noise of a conventional heterodyne detector and the well-known 3 dB extra quantum noise was not present in the experiment as predicted by theory. We carefully examined the imperfections of the experiment that could lead to the escape of the predicted noise from observation. The results highly suggest farther investigations for deeper understanding of the physics relevant to the quantum noise in optical heterodyne detection. 

\begin{acknowledgments}
This work was supported by the National Natural Science Foundation of China (grant No. 11174094).

\end{acknowledgments}

\bibliography{NoiselessPIA}

\end{document}